\documentclass[preprint,showpacs,preprintnumbers,amsmath,amssymb]{revtex4}

\usepackage{graphicx}
\usepackage{dcolumn}
\usepackage{bm}
\usepackage{color}

\begin{document}
\title{Overall current-voltage characteristics of space charge controlled currents for thin films by a single carrier species}
\author{
Kazuhiko Seki
}
\email{k-seki@aist.go.jp}
\affiliation{
National Institute of Advanced Industrial Science and Technology (AIST)\\
AIST Tsukuba Central 5, Higashi 1-1-1, Tsukuba, Ibaraki, Japan, 305-8565
}
\preprint{}
\begin{abstract}
The Mott-Gurney equation (Child's law) has been frequently applied 
to measure the mobility of carrier transport layers. 
One of the main assumption in the Mott-Gurney theory is ignoring the diffusive currents. 
It was not obvious, however, whether the diffusive currents can be ignored for thin carrier transport layers. 
We obtained the current-voltage relation using 
analytical solutions of drift-diffusion equation coupled with the Poisson's equation. 
The integration constants were numerically determined using  nonlinear equations obtained from boundary conditions.
A simple analytical relation between the voltage and current was also derived.   
The analytical equation improved over the Mott-Gurney equation 
when the voltage is between 0.1 and 2 [V] at room temperature. 
By using published data, we show that both the mobility and the layer thickness can be simultaneously obtained  
by applying the analytical expression. 
The effect of diffusion on the current-voltage relation  is explained by  
the movement of the virtual electrode formed by space charge accumulation. 

\end{abstract}

\maketitle
\newpage
\setcounter{equation}{0}
\section{Introduction}
\vspace{0.5cm}
Recently, thin carrier transport layers have been used for organic electronics 
such as electroluminescence and organic solar cells. \cite{Brutting}
Carrier transport in thin layers could be different from that in thick layers. 
Drift currents may dominate over the diffusive currents if the layer is sufficiently thick but it is not throughly understood  
whether the diffusive currents can be ignored or not for thin carrier transport layers.  
By ignoring the diffusive currents, 
a simple analytical expression describing the relation between the current and voltage has been derived. 
The simple equation called the Mott-Gurney equation (Child's law) has been widely applied to measure the mobility.  \cite{Brutting,Mott}
When the carrier injection into organic carrier transport layers is sufficiently fast, 
the carrier transport is limited by the space charge accumulated at the injected side. 
The Mott-Gurney equation was derived by approximately solving the drift-diffusion equation  coupled with Poisson's equation 
under the space charge limited condition. 
The coupled non-linear equations were solved by ignoring diffusive currents. 
However, the injected carriers may accumulate at the interface  
and they give rise to the diffusive currents 
flowing  into the counter electrode for thin charge transport layers. 

Historically, the effect of diffusion on the Mott-Gurney equation was studied for highly resistive inorganic semiconductors. \cite{Wright61,Sinharay64,Fan48,Bonham}
The effect of the carrier diffusion on the space-charge limited currents was studied by 
solving approximately the drift-diffusion equation coupled with Poisson's  equation. \cite{Wright61,Sinharay64}
The approximate analytical solution improved the  Mott-Gurney equation as the voltage was decreased. 
The equation, however, was complicated. \cite{Wright61}
The boundary between the resistive semiconductor and the counter electrode was regarded as blocking contact. \cite{Wright61,Sinharay64}
It was also shown that the effect of diffusion on the  Mott-Gurney equation is not influenced by the nature of the blocking electrode.  \cite{Wright61}
In early works, the formal solutions of the drift-diffusion equation coupled with Poisson's  equation were expressed in terms of the Bessel functions. \cite{Wright61,Skinner}
They were also expressed using Airy functions. \cite{Sinharay64,Fan48}
Although Airy functions can be equivalently expressed by the Bessel functions, 
different kinds of the Bessel functions should be used depending on the direction of electric field. 
In this sense, the boundary conditions can be set easier by using the Airy functions. 
Using the boundary conditions, 
the non-linear equations to determine the integration constants were obtained.  
Even though Airy functions were used,
the resultant equations were very complicated and hard to solve analytically. \cite{Wright61,Skinner}
Various approximations were introduced to obtain the current-voltage relations. \cite{Wright61,Sinharay64} 
Recently, a simple analytical model is proposed instead of solving non-linear equations 
for the integration constants 
to obtain a current-voltage relation at low voltages. \cite{Kim} 
Indeed, the current-voltage relation is largely affected by the boundary conditions at low voltages compared to that at high voltages. 

There has been revived interest on the exact solutions of the drift-diffusion equation coupled with Poisson's  equation 
expressed using  
Airy functions. \cite{Neumann,Genenko,Shashkin}
Airy functions were applied to solve numerically the expressions describing charge injection at boundaries and transport in insulating medium 
in a self-consistent manner.   \cite{Neumann,Genenko,Shashkin}
In the self-consistent approach, 
the boundary conditions were given for both the injection over the barrier at low voltages  and 
the formation of space charge at high voltages. 

In this paper, we study analytically the non-linear equations derived from the boundary conditions. 
The non-linear equations obtained from the boundary conditions were solved approximately in the space-charge limit. 
By systematically investigating  the boundary conditions for the space-charge,  
we obtained the approximate solution of the nonlinear drift-diffusion equation coupled with Poisson's equation.  
The space charge accumulated in the vicinity of the injection electrode gives rise to 
the electric field directing opposite to the current flow at the injection interface 
due to repulsive interaction among space charges. \cite{Kao} 
The electric field at the counter electrode is in the direction of the current flow and 
the virtual electrode can be defined for the plane where the electric field is zero. \cite{Kao}
We obtained analytical expressions characterizing length of space charge accumulation, electrostatic potential and electric fields inside 
the carrier transport layers. 

A simple approximate equation expressing the current-voltage relation was derived and tested against 
the numerical exact results and experimental results. 
The approximate expression generalized the Mott-Gurney equation by taking into account the diffusion effect. 
Compared to the previous current-voltage relation taking into account the diffusion effect, \cite{Wright61}
our expression is simpler and reproduces the numerical results for wide range of the variation in voltage. 
By using the analytical results, we show that the diffusion effect is different from 
Ohm's law although the currents can be phenomenologically fitted by assuming linear voltage-dependence. 
The deviation of the current-voltage relation from the Mott-Gurney relation can be interpreted as 
the diffusion effect to move the virtual electrode toward the counter electrode as the drift current is decreased. 
Although the diffusion effect was suggested, \cite{Kao}
we are able to examine it rigorously using analytical results. 

In Sec. \ref{sec:II}, the diffusion effect on the current-voltage relation is formulated. 
The method to calculate the space charge limited current is introduced in Sec. \ref{sec:SCL}. 
The space charge is characterized in Sec. \ref{sec:ChaSCL}. 
In Sec. \ref{sec:Exp}, 
theoretical results are applied to analyze experimental results. 
Summary and discussion are given in  Sec. \ref{sec:summary}. 
In the appendix A, derivation of analytical solution of the drift-diffusion equation  coupled with Poisson's equation is summarized. 
A method to determine the integration constants is given in the Appendix B. 
The solution when the virtual electrode is equal to the injection interface is given in the Appendix C for comparison. 
\section{Effect of diffusion on the space-charge-limited current}
\label{sec:II}
We consider the case that the charge transport layer is sandwiched between the two electrode. 
The electric field is applied in the direction perpendicular to the surface of the electrode. 
By using the $x$-coordinate parallel to the direction of the electric fields, 
we express that 
positive carriers are injected at $x=0$ and absorbed at $x=L$. 
Our interest is a steady state current $J$. 
The carrier concentration $n(x)$ obeys one-dimensional drift-diffusion equation given by, 
\begin{align}
J &= -D \left[ \frac{\partial}{\partial x} n (x) - \frac{e E(x)}{k_B T}  n (x) \right], 
\label{eq:J}
\end{align}
where $e$ denotes the charge of the carriers, $E(x)$ denotes the electric field, and 
$D$ is the diffusion coefficient of carriers. 
$k_B$ and $T$ represent the Boltzmann constant and temperature, respectively. 
The diffusion coefficient can be expressed in terms of the mobility $\mu$ by using the Einstein relation $D=\mu k_B T$. 
The electrical mobility is given by $e \mu$. 

The electric field obeys Gauss's law, 
\begin{align}
\epsilon \epsilon_0 \frac{\partial}{\partial x} E (x) = e n (x) ,  
\label{eq:Poisson}
\end{align}
where $\epsilon$ is the relative dielectric constant of the carrier transport layer and 
$\epsilon_0$ is the vacuum permittivity. 
When $n(x)$ is determined from Eq. (\ref{eq:J}),  
$E (x)$ can be determined from Eq. (\ref{eq:Poisson}). 
On the other hand, 
Eq. (\ref{eq:J}) can be solved once $E (x)$ is determined. 
In the below, $n(x)$ and $E (x)$ are determined in a self-consistent way 
to satisfy both Eqs. (\ref{eq:J}) and (\ref{eq:Poisson}) simultaneously. 

By substituting Eq. (\ref{eq:Poisson}) into Eq. (\ref{eq:J}), we obtain a closed equation for $E(x)$, 
\begin{align}
-e J = \epsilon \epsilon_0 D \frac{\partial}{\partial x} 
\left[ 
\frac{\partial}{\partial x} E (x) - 
\frac{e}{2 k_B T} E^2 (x) 
\right] . 
\label{eq:closedE}
\end{align}
Integration of Eq. (\ref{eq:closedE}) yields, 
\begin{align}
-eJ(x+L C_E) = \epsilon \epsilon_0 D
\left[ 
\frac{\partial}{\partial x} E (x) - 
\frac{e}{2 k_B T} E^2 (x) 
\right] , 
\label{eq:integrateclosedE}
\end{align}
where $C_E$ is a constant of integration. 
The solution is given in terms of 
a pair of linearly independent solution of the Airy equation, ${\rm Bi} (z)$ and ${\rm Ai} (z)$, \cite{Abramowitz}
and their derivatives as 
(\cite{Sinharay64,Fan48} see the Appendix A.)
\begin{align}
E(x)=-\frac{2 k_B T}{e} 
\left( \frac{2 \pi J r_c}{D} \right)^{1/3}
\frac{{\rm Bi}^\prime (z)  +  C_B\,  {\rm Ai}^\prime (z)}{{\rm Bi} (z) + C_B\,  {\rm Ai}  (z)}  ,  
\label{eq:solAiry}
\end{align}
where $z$ is given by, 
\begin{align}
z= z_L \left(\frac{x}{L}+C_E\right),  
\label{eq:z}
\end{align}
$z_L$ denotes the dimensionless parameter characterizing the space charge expressed by 
\begin{align}
z_L \equiv \left( \frac{2 \pi J r_c}{D} \right)^{1/3} L ,  
\label{eq:zL}
\end{align} 
the Onsager length (Coulomb radius) of the hole transport layer is given by 
\begin{align}
r_c=\frac{e^2}{4\pi \epsilon \epsilon_0 k_B T}, 
\label{eq:Onsagerl}
\end{align}
and $C_B$ is a constant of integration. 

By combining Eqs. (\ref{eq:Poisson}) and (\ref{eq:integrateclosedE}), the carrier density can be expressed as 
\begin{align}
n(x) = \frac{\epsilon \epsilon_0}{2 k_B T} E^2 (x) -\frac{J}{D}L \left(\frac{x}{L}+ C_E \right),  
\label{eq:comb}
\end{align}
where $E(x)$ is given by Eq. (\ref{eq:solAiry}). 
Equation (\ref{eq:comb}) is useful to express the boundary conditions given by $n(x)$ in terms of $E (x)$. 
The constants of integration, $C_B$ and $C_E$, can be determined from the boundary conditions. 
The quasi-Fermi energy $\phi_f (x)$ can be defined using $n(x)$ as 
$n(x)\sim \exp \left[- \phi_f (x) / (k_{\rm B} T) \right]$. 
From Eq. (\ref{eq:comb}), 
the quasi-Fermi energy is expressed using $E(x)$ given by Eq. (\ref{eq:solAiry}) as, 
\begin{align}
\phi_f (x)= - k_{\rm B} T \ln 
\left[ \frac{\epsilon \epsilon_0}{2 k_B T} E^2 (x) -\frac{J}{D}L \left(\frac{x}{L}+ C_E \right)
\right] . 
\label{eq:Fermi}
\end{align}

The applied voltage affects the electrostatic potential $\phi(x)$ satisfying 
$E= - \partial \phi (x)/(\partial x)$. 
By integration, the potential can be expressed as
\begin{align}
\phi (x) = 2 \frac{k_{\rm B} T}{e}  
\ln \left| {\rm Bi} (z) + C_B\,  {\rm Ai}  (z)
\right|   
\label{eq:pot}
\end{align}
apart from a constant. 
The applied voltage is related to the potential difference at the both boundaries, $V=\phi(0)-\phi(L)$ and 
can be written as 
\begin{align}
V= 2 \frac{k_{\rm B} T}{e} \ln \left|\frac{{\rm Bi} (z_L C_E) + C_B\,  {\rm Ai}  (z_L C_E)}{{\rm Bi} \left[z_L(1+C_E)\right] + C_B\,  {\rm Ai} \left[z_L(1+C_E)\right]}
\right| .
\label{eq:V_formal}
\end{align}

\section{Space charge limited current}
\label{sec:SCL}
The voltage given by Eq. (\ref{eq:V_formal}) is related to the current through $z_L$ defined by Eq. (\ref{eq:z}). 
The relation between the current and voltage is non-linear and can vary according to the values of 
the integration constants, $C_E$ and $C_B$. 
The distribution of carriers and the profile of electric fields are nonlinear function of the distance from 
the carrier injection contact interface. 
The integration constants, $C_E$ and $C_B$, can be obtained from the boundary conditions set at 
the injection and collection contact interfaces.  
In order to consider the boundary conditions, 
it is convenient to introduce the dimensionless density given by 
$\bar{n}(X)\equiv n (x) 4 \pi r_c L^2$  with $X=x/L$ and express  
Eq.  (\ref{eq:comb}) in terms of $z=z_L(X+C_E)$ as 
\begin{align}
\frac{\bar{n}(X)}{2 z_L^2}= - z_L \left(X+ C_E\right)+ \left( 
\frac{Bi^\prime (z)  + Ai^\prime (z) \,  C_B}{Bi (z) + Ai  (z)\, C_B}
\right)^2 ,
\label{eq:sol14}
\end{align}
where $J_{dl}= 2 z_L^3$ and $z_L=\left[2 \pi J r_c/D \right]^{1/3} L$ are used. 
We also introduce  the quantity proportional to the dimensionless electric field defined by $\bar{E} = e E(x) L/(k_B T)$, 
\begin{align}
f(z) =  -\frac{\bar{E}}{2 z_L} =   \frac{Bi^\prime (z)  + Ai^\prime (z) \,  C_B}{Bi (z) + Ai  (z)\, C_B}  . 
\label{eq:nonE}
\end{align}
By setting the boundary conditions on $\bar{n}(0)$ and $\bar{n}(1)$, 
the integration constants can be determined through $f(z_0)$ and $f(z_1)$, 
where $z_0$ and $z_1$ are given in terms of $z_L$ defined by Eq. (\ref{eq:zL}) as 
\begin{align}
z_0= z_L C_E\mbox{ and } z_1= z_L \left(1+C_E\right) .
\label{eq:z0z1}
\end{align} 

We assume the limit of fast injection from the electrode to the carrier transport layer. 
As a result, 
carriers are  accumulated in the carrier transport layer by the injection from the electrode  
and the contact can be regarded as a charge reservoir for the carrier transport layer. \cite{Hertel}
The excess mobile carriers is referred to as the space charge. 
The electric field is reduced to 
zero to resolve the carrier accumulation 
by the mobile carriers. \cite{Hertel} 
The forward bias of the electric field applies to carriers when the distance from the injection interface is larger than 
that of the location where the electric field becomes zero. 
In this sense, the location where the electric field becomes zero can be regarded as the virtual electrode. \cite{Kao}
In this section, we consider the case that the virtual electrode presents in the carrier transport layer. 
The situation can be stated that  
$f(z)$ crosses the z-axis at some point between $z_0$ and $z_1$. 
For positive carriers 
the electric field at the collection interface is positive. 
The electric field changes sign by decreasing the distance from the injection interface.  
The electric field at the injection interface is assumed to be negative. 
Corresponding to the change in the sign of the electric fields and according to Eq. (\ref{eq:nonE}), 
$f(z)$ is negative when $z=z_1$ and positive when $z=z_0$.

In the limit of fast injection from the electrode to the carrier transport layer, 
we set the boundary condition at the injection interface as, 
\begin{align}
n(0) &=n_0,  
\label{eq:BC1}
\end{align}
where $n_0$ is the carrier site density which could be occupied by injected carriers. 
By introducing the typical value, $n_0= 4.0 \times 10^{25}$ [m$^{-3}$], 
the dimensionless density at the interface is estimated as $\bar{n} (0) =4 \pi n_0 r_c L^2= 8.2 \times 10^4$ when 
the thickness of the carrier transport layer is $L=100$ [nm] and $\epsilon=3.5$. 
$\bar{n} (0) =4 \pi n_0 r_c L^2= 8.2 \times 10^6$ is estimated when $L=1$ [$\mu$m]. 
From the boundary condition and using Eq. (\ref{eq:sol14}), we obtain, 
\begin{align}
\sqrt{\frac{\bar{n}(0)}{2 z_L^2}+ z_0}= f (z_0), 
\label{eq:BCz0}
\end{align}
where $f(z)$ and $z_0$ are given by Eqs. (\ref{eq:nonE})-(\ref{eq:z0z1}), respectively.

When the electric field becomes zero between the injection interface and the counter electrode, 
the current is limited by the flow of the accumulated carriers from the virtual electrode. 
In this case, 
$C_B\gg 1$ is satisfied and $C_E$ is approximately given by  
\begin{align}
C_E \approx \frac{1}{z_L} \left( a_1 + 
\frac{1}{\sqrt{\bar{n}(0)/(2 z_L^2)+ a_1}}
\right). 
\label{eq:CE_approx}
\end{align}
(see the Appendix B for the determination of integration constants). 

The distance $x^\dagger$ of the virtual electrode from the injection interface can be approximately found from, 
$Ai^\prime [z_L(X^\dagger + C_E)]$
where $X^\dagger=x^\dagger/L$ denotes the dimensionless distance of the virtual electrode from the injection interface. 
By denoting the first zero of $Ai^\prime (z)$ on the negative $z$-axis by $a'(1)$, we obtain, 
$z_L(X^\dagger + C_E)= a_1'=-1.018 \cdots $. 
By introducing Eq. (\ref{eq:CE_approx}), $X^\dagger$ can be expressed as, 
\begin{align}
X^\dagger \approx \frac{1}{z_L} 
\left( - a_1+ a_1'- 
\frac{1}{\sqrt{\bar{n}(0)/(2 z_L^2)+ a_1}}
\right) . 
\label{eq:virtuale}
\end{align}
In the limit of $\bar{n}(0)/(2 z_L^2) \geq 100$, 
which can be satisfied in the space charge limit, 
the location of the virtual electrode can be approximately given by 
$X^\dagger \approx 1/z_L$ and 
\begin{align}
x^\dagger = \left(\frac{D}{2 \pi J r_c} \right)^{1/3} . 
\label{eq:virtuale1}
\end{align}
The location of the virtual electrode decreases by increasing the 
current obeying the power law with the exponent $-1/3$. 
The exponent was pointed out previously. \cite{Wright61}. 

By taking the limit of $C_B \rightarrow \infty$, Eq. (\ref{eq:V_formal}) simplifies into, 
\begin{align}
V= 2 \frac{k_{\rm B} T}{e} \ln \left|\frac{{\rm Ai}  (z_L C_E)}{{\rm Ai} \left[z_L(C_E+ 1)\right]}
\right| . 
\label{eq:V_formal_limit}
\end{align}
By substituting $C_E$ approximated by Eq. (\ref{eq:CE_approx}) into Eq. (\ref{eq:V_formal_limit}),  
$V$ is obtained as a function of $z_L$ expressed in terms of the current by Eq. (\ref{eq:zL}). 
The relation between the current and voltage given in terms of the Airy function can be further simplified. 
By noting that $z_L C_E$ is close to the first zero of ${\rm Ai}  (z)$ as shown by Eq. (\ref{eq:CE_approx}), 
we expand ${\rm Ai}  (z)$ around $a_1$, 
\begin{align}
{\rm Ai}  (z_L C_E)
= {\rm Ai}' (a_1) \frac{1}{\sqrt{\bar{n}(0)/(2 z_L^2)+ a_1}} . 
\label{eq:apprAi}
\end{align}
By substituting Eq. (\ref{eq:apprAi}) and introducing \cite{Abramowitz}
\begin{align}
{\rm Ai} (z) \sim \frac{\exp\left( - (2/3) z^{3/2}\right)}{2\sqrt{\pi}z^{1/4}} , 
\end{align}
obtained by taking the limit of $z \gg 1$, 
Eq. (\ref{eq:V_formal_limit}) can be written using Eq. (\ref{eq:CE_approx}) as, 
\begin{align}
V \approx \frac{k_{\rm B} T}{e} \left[ 
\frac{4}{3} z_L^{3/2}+2
\left( a_1+ \frac{1}{\sqrt{\bar{n}(0)/(2 z_L^2)+ a_1}} \right) \sqrt{z_L} + 
\ln \left( \frac{4 \pi \sqrt{z_L} {\rm Ai}' (a_1)^2}{a_1 + \bar{n}(0)/(2 z_L^2)} \right)
 \right]
, 
\label{eq:V_asympt}
\end{align}
where $a_1=-2.34 \cdots$ is the first zero of ${\rm Ai} (z)$ from the origin on the negative axis.  

Except for the logarithmic term, the leading terms can be rewritten as
\begin{align}
V \approx \frac{2}{3} \left( 
\frac{2I}{\epsilon \epsilon_0 e\mu}
\right)^{1/2} L^{3/2} +2
\left( a_1+ \frac{1}{\sqrt{\bar{n}(0)/(2 z_L^2)+ a_1}} \right) \left(
\frac{k_{\rm B} T}{e}
\right)^{2/3} 
\left( \frac{I}{2\epsilon \epsilon_0 e\mu}
\right)^{1/6} L^{1/2}  ,
\label{eq:V_asympt1}
\end{align}
where the electric current density is denoted by $I=e J$, $e \mu$ is the electrical mobility, 
and the mobility $\mu$ is introduced by using the Einstein relation $D=\mu k_{\rm B} T$. 
Retaining only the first term on the right-hand side yields the Mott--Gurney equation (Child's law) given by \cite{Mott}
\begin{align}
I = \frac{9}{8} e\mu \epsilon \epsilon_0 \frac{V^2}{L^3} . 
\label{eq:Child}
\end{align}
When $\bar{n}(0)/(2 z_L^2)>> |a_1|$, Eq. (\ref{eq:V_asympt1}) can be approximated as, 
\begin{align}
V \approx \frac{2}{3} \left( 
\frac{2I}{\epsilon \epsilon_0 e\mu}
\right)^{1/2} L^{3/2} -4.68 \left(
\frac{k_{\rm B} T}{e}
\right)^{2/3} 
\left( \frac{I}{2\epsilon \epsilon_0 e\mu}
\right)^{1/6} L^{1/2}  ,
\label{eq:V_resultasympt2}
\end{align}
where $a_1=-2.34$ is substituted. 
The first term leading to the Mott-Gurney equation (Child's law) is given in terms of the mobility while 
the second term depends on the diffusion coefficient besides the mobility by introducing $k_{\rm B} T=D/\mu$. 
In this sense, the second term represents the diffusion effect on the Mott-Gurney equation. 

Equation (\ref{eq:V_resultasympt2}) is independent of the boundary condition at the counter electrode. 
The reason is the following. 
As shown in the Appendix B, the integration constant $C_E$ is insensitive to the boundary condition at the counter electrode 
if $C_B \gg 1$, while  
the integration constant $C_B$ is essentially determined by both boundary conditions, {\it i.e.} 
the boundary condition at the injection interface and that at the counter electrode. 
By using reasonable values of the dimensionless extraction rate denoted by $\bar{k}_e$ and (or) 
the carrier concentration $n(L)$, 
we show $C_B \gg 1$ in the Appendix B and it will be verified using Fig. \ref{fig:C_E_C_B} when $2 z_L^3 \geq 100$. 
(see also Ref. \cite{Wright61} and the discussion following Eq. (\ref{eq:BCz1_r}))

\begin{figure}
\centerline{
\includegraphics[width=0.6\columnwidth]{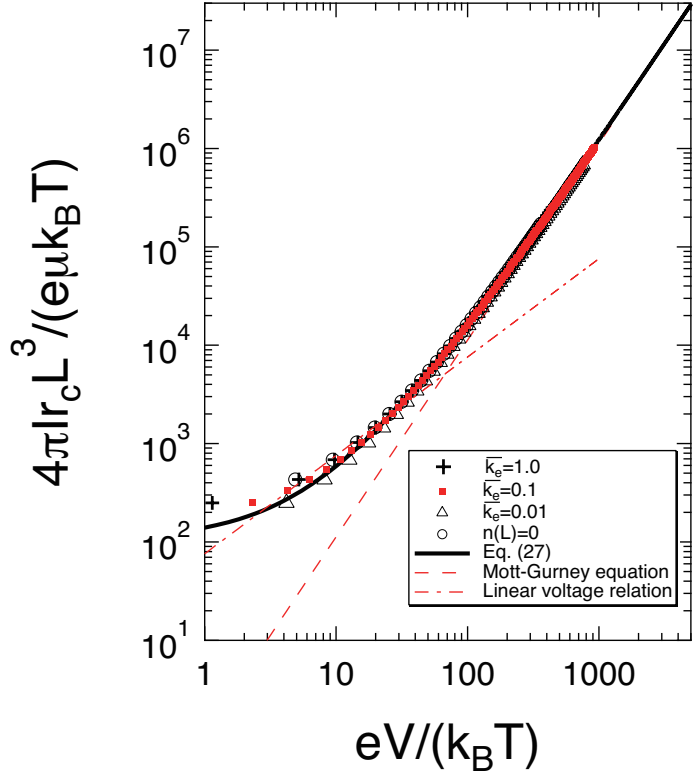}
}
\caption{(Color online) 
Dimensionless current  
$4 \pi I r_c L^3/(e \mu k_{\rm B} T)$ as a function of 
dimensionless voltage $eV/(k_{\rm B} T)$ for $\bar{n} (0) = 8.2 \times 10^4$. 
The crosses, (red) squares, triangles represent $\bar{k}_e=1$, $0.1$, and $0.01$ obtained from 
Eq. (\ref{eq:V_formal}) by numerically evaluating the boundary conditions,  
Eqs. (\ref{eq:BCz0}) and (\ref{eq:BCz1})  (see the text). 
The circles represent the results of $n(L)=0$. 
The  
thick solid line and (red) dashed line represent the results of Eq. (\ref{eq:V_resultasympt2}), 
and the Mott--Gurney equation, Eq. (\ref{eq:Child}), respectively.
The (red) dash-dot line indicates the fit by assuming a linear voltage dependence.}
\label{fig:SCL_IV_8200}
\end{figure}

In Fig. \ref{fig:SCL_IV_8200}, 
we compare the results of approximate analytical expressions to  
the numerically exact I-V characteristics. 
The exact I-V characteristics was obtained from numerically evaluating Eqs. (\ref{eq:BCz0}) and (\ref{eq:BCz1}) by
the Newton methods using the seeds for the integration constants 
obtained from Eqs. (\ref{eq:CE_approx}) and (\ref{eq:C_B}). 
The numerical calculations were performed using Mathematica. \cite{Mathematica}
We have used $\bar{n} (0) = 8.2 \times 10^4$. 
Essentially the same results can be found when $\bar{n} (0) = 8.2 \times 10^2$ (not shown). 

In Fig. \ref{fig:SCL_IV_8200}, the dash-dot line indicates the fit
by assuming a linear voltage dependence using the numerical results of $\bar{k}_e=0.1$ below 
$eV/(k_{\rm B} T)=100$. 
The obtained fitting function is given by $2 z_L^3 \approx 96 eV/(k_{\rm B} T)$. 
The onset voltage 
characterizing the deviation from the Mott-Gurney equation by the diffusion effect at low voltage can be read from Fig. \ref{fig:SCL_IV_8200} 
as  $eV/(k_{\rm B} T)=80$; 
the corresponding voltage is below $2.0$ [V]. 
The fitting is phenomenological and the actual  diffusion effect is given by 
the second term in Eq. (\ref{eq:V_resultasympt2}). 
As shown in Fig. \ref{fig:SCL_IV_8200}, Eq. (\ref{eq:V_resultasympt2}) captures the diffusion effect 
regardless of the boundary condition at the counter electrode.

\section{Characterization of Space charge}
\label{sec:ChaSCL}

\begin{figure}
\centerline{
\includegraphics[width=0.6\columnwidth]{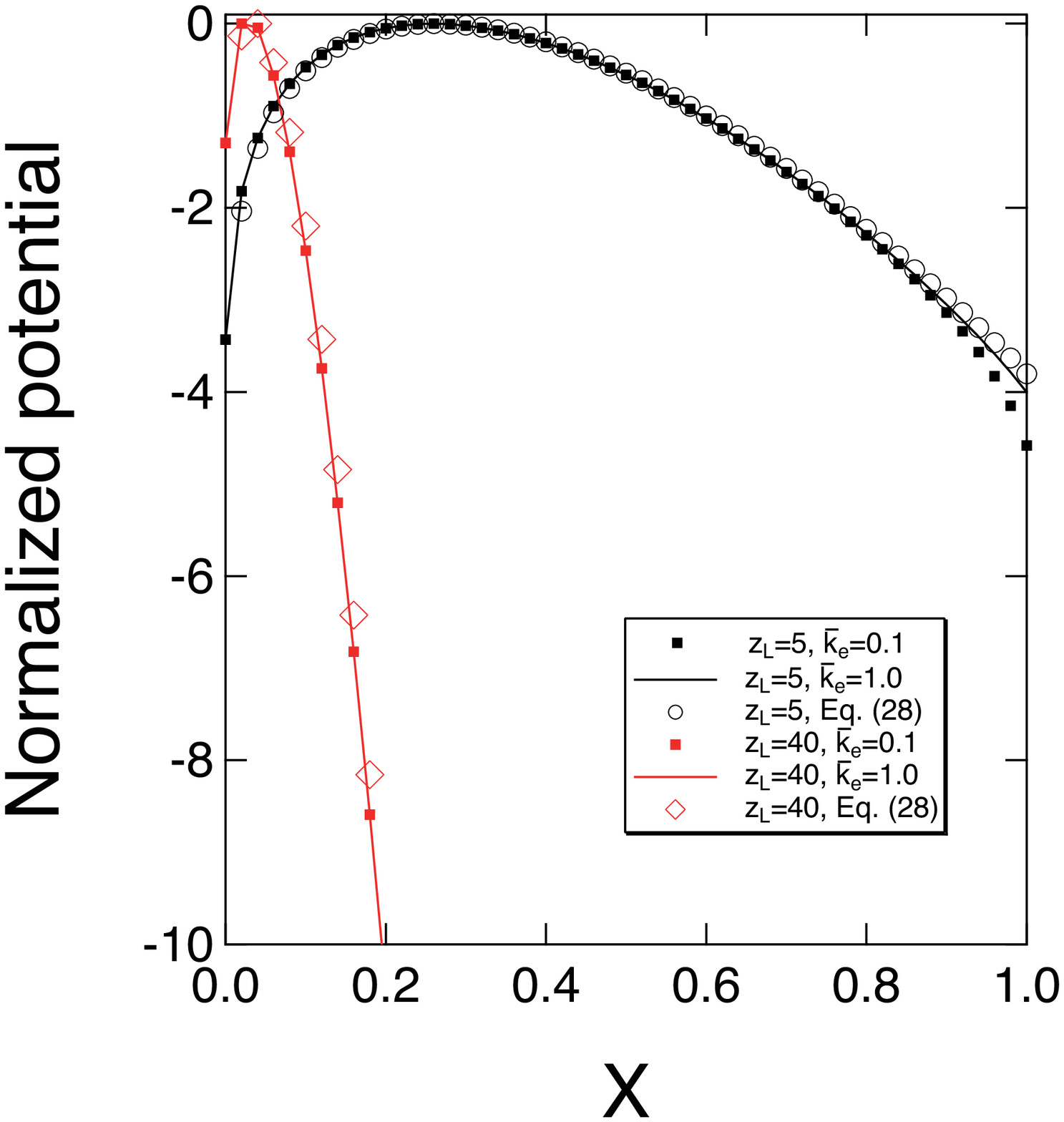}
}
\caption{(Color online) 
Potential  $\phi (x)$ normalized by the absolute maximum value as a function of 
the normalized distance from the injection interface denoted by $X=x/L$ for $\bar{n} (0) = 8.2 \times 10^4$. 
The upper (black) dots represent $z_L=5$ and $\bar{k}_e=0.1$. 
The upper (black) line denotes $z_L=5$ and $\bar{k}_e=1.0$. 
The circles indicate the results of Eq. (\ref{eq:pot_approx}) for $z_L=5$. 
The lower (red) dots represent $z_L=40$ and $\bar{k}_e=0.1$. 
The lower (red) line denotes $z_L=40$ and $\bar{k}_e=1.0$. 
The (red) diamonds indicate the results of Eq. (\ref{eq:pot_approx})  for $z_L=40$. 
}
\label{fig:potential_X}
\end{figure}

\begin{figure}
\centerline{
\includegraphics[width=0.6\columnwidth]{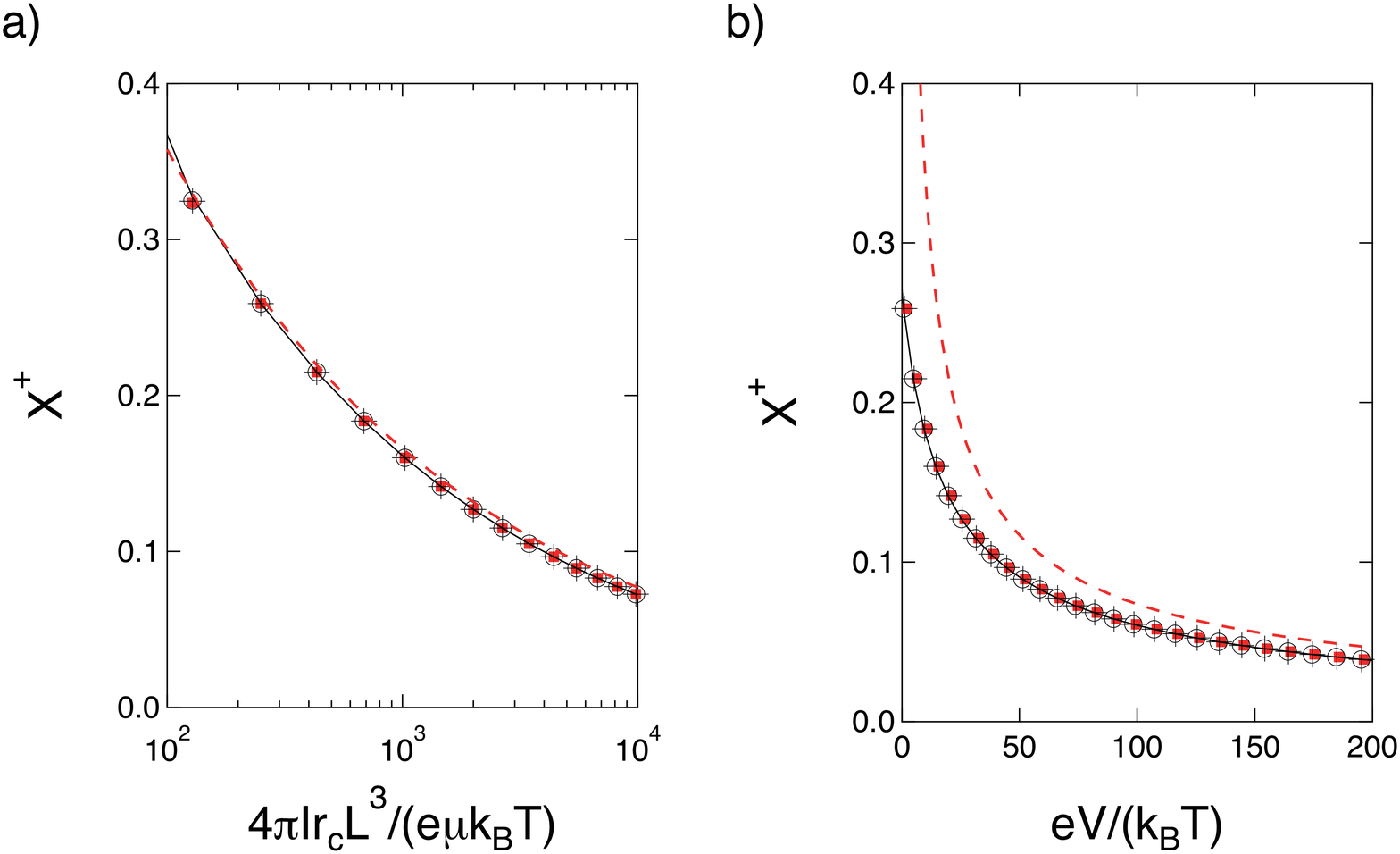}
}
\caption{(Color online) 
$X^\dagger=x^\dagger/L$ 
as a function of a) dimensionless current  
$4 \pi I r_c L^3/(e \mu k_{\rm B} T)$ and 
b) dimensionless voltage $eV/(k_{\rm B} T)$ for $\bar{n} (0) = 8.2 \times 10^4$. 
The crosses and (red) dots represent $\bar{k}_e=1.0$ and $0.1$. 
The circles represent the results of $n(L)=0$. 
The thin solid line represent $X^\dagger$ approximately calculated from 
$z_L(X^\dagger + C_E)= a_1'$ when $n(L)=0$.
The (red) dashed lines in the left and right figures represent the results of Eq. (\ref{eq:virtuale1nd}), 
and Eq. (\ref{eq:virtuale2}), respectively.}
\label{fig:SCL_virtual_electrode}
\end{figure}

When the space-charge is formed by the injection of carriers, 
the electrostatic potential increases by the accumulated carriers.  
The location of the maximum potential roughly indicates the region where the accumulated carriers are started to flow driven by the electric field. 
The electrostatic potential $\phi (x)$ normalized by the absolute maximum value  is shown as a function of $X=x/L$ for $\bar{n} (0) = 8.2 \times 10^4$ 
in Fig. \ref{fig:potential_X}. 
The location of the potential maximum shifts to the left by increasing the 
currents as it will be shown in Fig. \ref{fig:SCL_virtual_electrode}. 
By using Eq. (\ref{eq:CE_approx}) and $|C_B | \gg 1$, 
Eq. (\ref{eq:pot}) can be approximated as, 
\begin{align}
\phi (x) = 2 \frac{k_{\rm B} T}{e}  
\ln \left| {\rm Ai}  \left[\left( \frac{2 \pi J r_c}{D} \right)^{1/3}x + a_1 \right]
\right| , 
\label{eq:pot_approx}
\end{align}
where $a_1=-2.34 \cdots$ is the first zero of ${\rm Ai} (z)$ and $r_c$ is the Onsager length given by Eq. (\ref{eq:Onsagerl}). 
The overall potential can be well approximated by Eq. (\ref{eq:pot_approx}) except in the vicinity of 
the counter electrode.

\begin{figure}
\centerline{
\includegraphics[width=0.6\columnwidth]{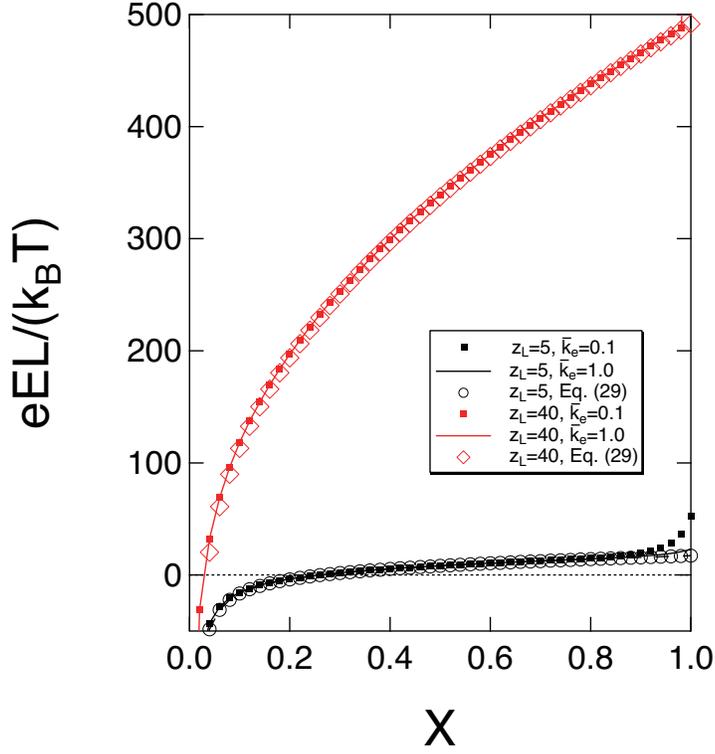}
}
\caption{(Color online) 
Dimensionless electric fields $eEL/(k_{\rm B} T)$ as a function of 
the normalized distance from the injection interface denoted by $X=x/L$ for $\bar{n} (0) = 8.2 \times 10^4$. 
The lower (black) dots represent $z_L=5$ and $\bar{k}_e=0.1$. 
The lower (black) line denotes $z_L=5$ and $\bar{k}_e=1.0$. 
The circles indicate the results of Eq. (\ref{eq:solAiry_approx}) for $z_L=5$. 
The upper (red) dots represent $z_L=40$ and $\bar{k}_e=0.1$. 
The upper (red) line denotes $z_L=40$ and $\bar{k}_e=1.0$. 
The (red) diamonds indicate the results of Eq. (\ref{eq:solAiry_approx}) for $z_L=40$. 
The dotted line denotes the line of $E=0$.   
}
\label{fig:efields}
\end{figure}

The electric fields as a function of the distance from the injection interface are shown in Fig. \ref{fig:efields}. 
The direction of the electric fields in the vicinity of the injection interface is opposite to that of carrier flow. 
The virtual electrode can be defined at the distance when the electric fields become zero.  \cite{Kao}
The drift carrier flow to the counter electrode occurs from 
the virtual electrode where  
the electric fields are zero. \cite{Kao}
The drift flow to the counter electrode is supplied by the carriers 
accumulated between the virtual electrode and the injection interface. 
Carriers accumulated between the injection interface and the virtual electrode can be regarded as 
a carrier reservoir for the current flow to the counter electrode. 

The electric fields are non-linear function of the distance from the injection interface and  
the increasing rates decrease by increasing the distance. 
The nonlinear growth of the electric fields is caused by the inhomogeneous carrier distribution by the accumulated carriers. 
The distance dependence is approximately expressed using the same approximation leading to Eq. (\ref{eq:pot_approx}) as
\begin{align}
E(x)=-\frac{2 k_B T}{e} \left( \frac{2 \pi J r_c}{D} \right)^{1/3}
\frac{{\rm Ai}^\prime \left[ \left(2\pi J r_c/D \right)^{1/3}x + a_1\right]}{{\rm Ai} \left[ \left(2 \pi J r_c/D \right)^{1/3}x + a_1\right]} .
\label{eq:solAiry_approx}
\end{align}
When the extraction rate is decreased from $\bar{k}_e=1.0$ to $\bar{k}_e=0.1$ 
by keeping $z_L=5$ unaltered, 
the electric fields in the vicinity of the counter electrode are changed as shown in Fig. \ref{fig:efields}. 
The region affected by the extraction rate is localized in the vicinity of the counter electrode. 
The electric fields in the other regions  are not affected by $\bar{k}_e$. 
In addition, 
if the carrier distribution except the vicinity of the counter electrode is not affected by $\bar{k}_e$,   
the currents are not affected by $\bar{k}_e$.  

\begin{figure}
\centerline{
\includegraphics[width=0.6\columnwidth]{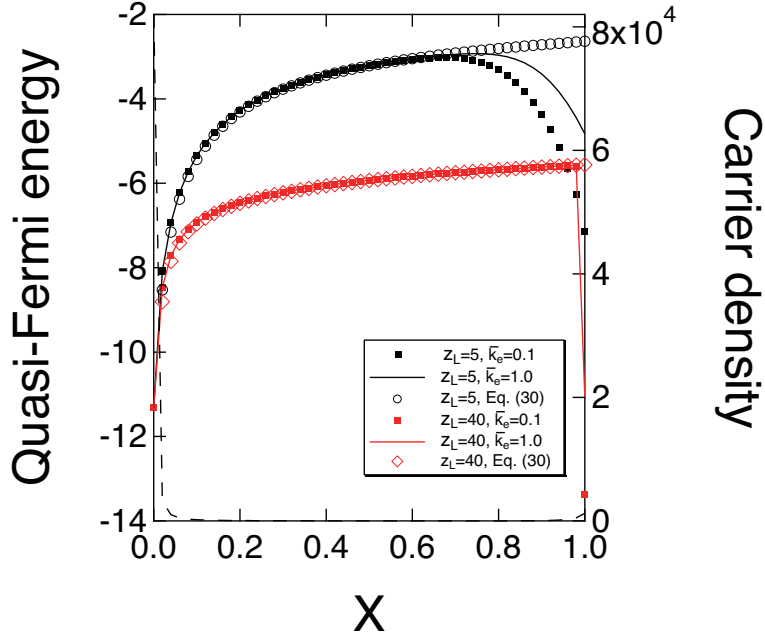}
}
\caption{(Color online) 
Quasi-Fermi energy (arbitrary unit) as a function of 
the normalized distance from the injection interface denoted by $X=x/L$ for $\bar{n} (0) = 8.2 \times 10^4$. 
The right axis is the normalized carrier density as a function of $X$. 
The upper (black) dots represent $z_L=5$ and $\bar{k}_e=0.1$. 
The upper (black) line denotes $z_L=5$ and $\bar{k}_e=1.0$. 
The circles indicate the approximate expression (see the text) for $z_L=5$. 
The lower (red) dots represent $z_L=40$ and $\bar{k}_e=0.1$. 
The lower (red) line denotes $z_L=40$ and $\bar{k}_e=1.0$. 
The (red) diamonds indicate the approximate expression for $z_L=40$ (see the text). 
The dashed line denotes the carrier density for $z_L=5$ and $\bar{k}_e=0.1$.   
}
\label{fig:quasiFermi_X}
\end{figure}

The carrier distribution and the resultant quasi-Fermi energy 
 $\phi_f (x)$ defined using $n(x)$ as 
$n(x)\sim \exp \left[- \phi_f (x) / (k_{\rm B} T) \right]$ are shown in Fig. \ref{fig:quasiFermi_X}. 
The quasi-Fermi energy is not homogeneous and reflects  
the carrier accumulation by the carrier injection. 
When the extraction rate is decreased from $\bar{k}_e=1.0$ to $\bar{k}_e=0.1$, 
the quasi-Fermi energy is affected only in the vicinity of the counter electrode. 
The other region can be well approximated by the quasi-Fermi energy obtained using,  
\begin{align}
n(x)=- \frac{J}{D} x + 
\left( \frac{J^2}{2 \pi r_c D^2} 
\right)^{1/3}
\left\{-a_1 + 
\frac{{\rm Ai}^\prime \left[ \left(2 \pi J r_c/D \right)^{1/3}x + a_1\right]}{{\rm Ai} \left[ \left(2 \pi J r_c/D \right)^{1/3}x + a_1\right]} 
\right\} .   
\label{eq:population_approx}
\end{align}

The location of the virtual electrode shifts according to the strength of currents as shown in Fig. \ref{fig:SCL_virtual_electrode}. 
The virtual electrode moves toward the counter electrode by decreasing the currents. 
In Fig. \ref{fig:SCL_virtual_electrode}, 
we show the location of the virtual electrode determined numerically from $f(z)=0$ using Eq. (\ref{eq:nonE}). \cite{Kao}
We also calculated $X^\dagger$ from 
$z_L(X^\dagger + C_E)= a_1'$ and found that the difference is negligibly small. 
The dashed  line in Fig. \ref{fig:SCL_virtual_electrode} a) represents the results of  
\begin{align}
X^\dagger \approx \frac{1}{z_L} 
\left( - a_1+ a_1' \right)\approx \frac{1.32}{z_L} . 
\label{eq:virtuale1nd}
\end{align}
Equation (\ref{eq:virtuale1nd}) can be expressed as, 
\begin{align}
x^\dagger \approx 1.32 \left( \frac{D}{2 \pi J r_c} \right)^{1/3} . 
\label{eq:virtuale1_1}
\end{align}
By assuming the Mott-Gurney equation, Eq. (\ref{eq:Child}), 
Eq. (\ref{eq:virtuale1nd}) can be rewritten as, 
\begin{align}
X^\dagger \approx \left(2 \frac{k_{\rm B} T}{eV}\right)^{2/3} .
\label{eq:virtuale2}
\end{align}
The results of Eq. (\ref{eq:virtuale2}) are shown as (red) dashed line in Fig. \ref{fig:SCL_virtual_electrode} b). 
As shown in the figure, the results overestimate the numerical results in particular when $eV/(k_{\rm B} T)$ is smaller than $80$. 
The large deviation originates from the diffusion effect on the Mott-Gurney equation.  
According to Fig. \ref{fig:SCL_virtual_electrode},  
the Mott-Gurney equation is applicable when the voltage is large enough so that 
the distance of the virtual electrode from the injection interface is within $10$ \%  of the total width of the 
carrier transport layer. 
Even when the voltage is low, 
the location of the virtual electrode given by Eq. (\ref{eq:virtuale1_1}) takes into account the diffusion effect and 
reproduces that obtained using the exact numerical calculation.  

\begin{figure}
\centerline{
\includegraphics[width=0.6\columnwidth]{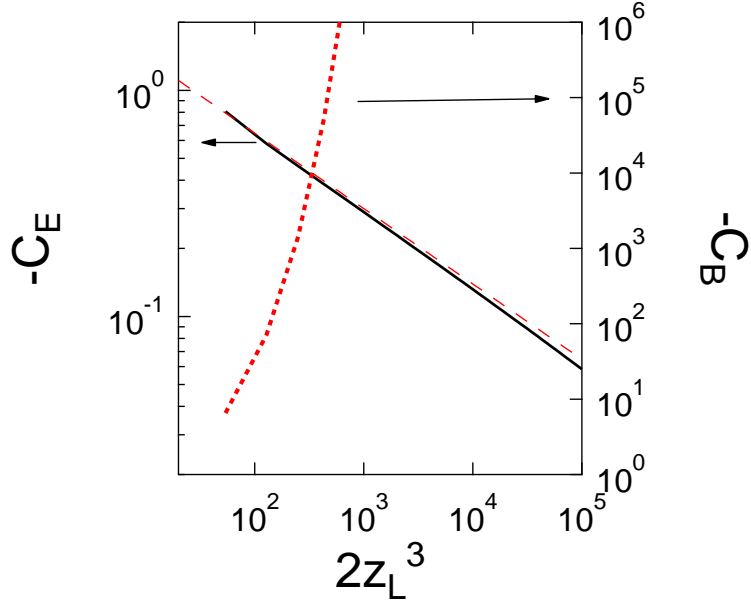}
}
\caption{(Color online) 
$-C_E$ and $-C_B$ as a function of $2 z_L^3$ for $\bar{n} (0) = 8.2 \times 10^4$ and $\bar{k}_e=0.1$. 
The thick solid line represents $-C_E$. 
The red dots represent $-C_B$. 
The (red) dashed line represents the result of $C_E \sim a_1/z_L$. }
\label{fig:C_E_C_B}
\end{figure}

In Fig. \ref{fig:C_E_C_B}, we show $-C_E$ and $-C_B$ as a function of $2 z_L^3$ for $\bar{n} (0) = 8.2 \times 10^4$ and 
$\bar{k}_e=0.1$. 
We note that $-C_B \gg 1$ is satisfied and $-C_B$ increases rapidly as $2 z_L^3$ increases. 
Using Eq. (\ref{eq:CE_approx}) and $\bar{n} (0) \gg 2 a_1 z_L^2$, 
we obtain $C_E \sim a_1/z_L$. 
The line of $C_E \sim a_1/z_L$ is close to the numerical results of $C_E$. 
The approximate current-voltage relation, Eq. (\ref{eq:V_resultasympt2}), was derived under the condition of 
$|C_B| \gg 1$. 
The condition is satisfied when $2 z_L^3 > 100$ as shown in Fig. \ref{fig:C_E_C_B}. 
The condition is consistent with the results shown in Fig. \ref{fig:SCL_IV_8200};  
equation (\ref{eq:V_resultasympt2}) reproduces the exact numerical results when $eV/(k_{\rm B} T)\geq 5$ and $2 z_L^3 \geq 500$. 

As shown in the Appendix C, 
the current-voltage relation obeys the Mott-Gurney equation to the low voltage given by $k_B T/e=0.026$ [V] 
if the boundary condition at the injection interface is given by $E(0)=0$;   
the boundary condition indicates that 
the virtual electrode is equal to the injection interface.  
The boundary condition is unrealistic at low voltages where 
the virtual electrode is moved away from the injection interface. 
In Fig. \ref{fig:SCL_IV_8200},  
the deviation from the Mott-Gurney equation occurs at low voltages below the onset voltage. 
The onset voltage is much higher than $k_B T/e=0.026$ [V] and  
the deviation correlates with the diffusion effect to move the virtual electrode toward the counter electrode as shown in 
Fig. \ref{fig:potential_X}. 
In Eq. (\ref{eq:V_resultasympt2}), 
the second term appeared by using the integration constant $C_E$ obtained from Eq. (\ref{eq:CE_approx}). 
The integration constant $C_E$ is determined from the boundary condition at the injection interface 
and is related to the formation of accumulated charges. 

In Eqs. (\ref{eq:comb})-(\ref{eq:pot}), 
the positional dependence is given in terms of $z_L x/L$ alone. 
The rapid changes in the density in the vicinity of either electrode 
shown in Fig. \ref{fig:quasiFermi_X} can be characterized by $L/z_L$. 
By defining the region of the charge accumulation for $x/L$ as $1/z_L$, 
we obtain $0.2$ and $0.025$ for $z_L=5$ and $40$, respectively. 
As shown in Fig. \ref{fig:quasiFermi_X}, the charge accumulation length is consistent with 
the profile of the quasi-Fermi energy obtained from the density profile. 
The charge accumulation length also characterizes the location of the virtual electrode given by Eq. (\ref{eq:virtuale1nd}). 
The charge accumulation length, $\ell_{\rm ca}$ can be defined as 
\begin{align}
\ell_{\rm ca}=\frac{L}{z_L} =\left(\frac{e \mu k_{\rm B} T}{2 \pi I r_c}\right)^{1/3}=\left( 
\frac{2 \mu \epsilon \epsilon_0 }{I e}
\right)^{1/3}\left( k_{\rm B} T\right)^{2/3},  
\label{eq:cacumulation}
\end{align}
 in the dimension form. 
 The charge accumulation length scales with $(\mu/I)^{1/3}$. 
 When the mobility is not altered, the accumulation length decreases by increasing the current density. 
For the same current density, the accumulation length increases according to a power law with the exponent $1/3$ as  the mobility increases. 

\section{Comparison with experiments}
\label{sec:Exp}

The space-charge limited currents in organic thin films were measured for a wide range of voltage to study 
layer thickness and temperature dependence 
in Au/alpha-conjugated sexithienyl/Au sandwich structures in Ref. \cite{Horowitz90}. 
The currents were proportional to the square of voltage as obtained from the Mott-Gurney equation at high voltages. 
By fitting the low voltage currents by assuming a linear voltage dependence, 
the conductivity was found to depend on the layer thickness in thin layers less than 2 $\mu$m. 
The conductivity of thick layers were found to be independent of layer thickness and we will not analyze the sample thicker than 2 $\mu$m. 
In experiments, there may be traps in the samples. 
Unfortunately, 
our analytical approach can be applied only when traps are shallow. 
When shallow trap states present and the traps are locally equilibrated with the free charge 
whose density is given by $n_f(x)$, 
the above results should be modified by introducing the substitution, 
$\mu \rightarrow \mu \theta$, 
where $\theta=n_f(x)/n(x)$.  
the total density $n(x)$ is the sum of  $n_f(x)$  to the trapped carrier density. 
In the below, the mobility may include the factor $\theta$.  

\begin{figure}
\centerline{
\includegraphics[width=0.6\columnwidth]{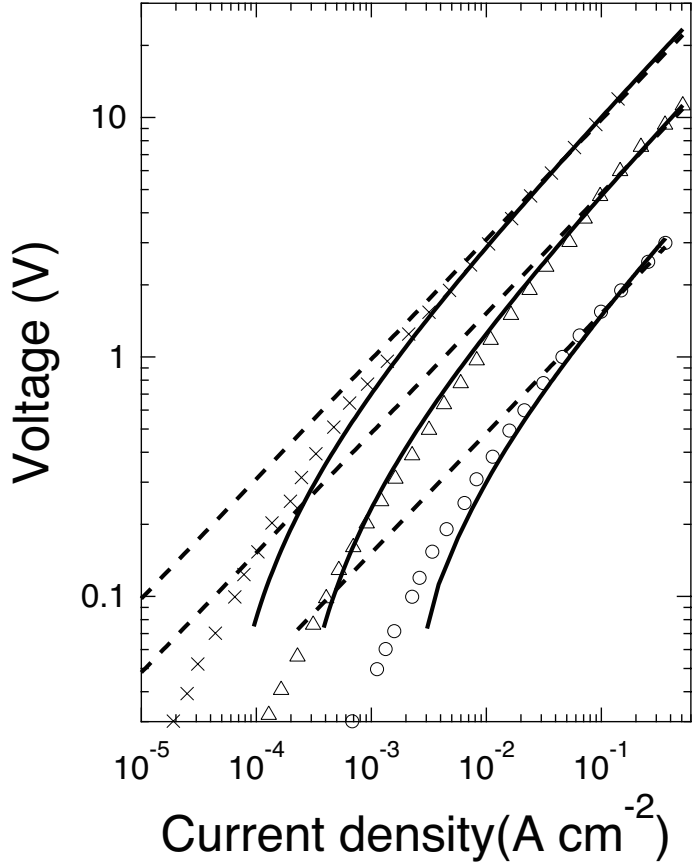}
}
\caption{The current-voltage relation of Au/$\alpha$-sexithienyl/Au structures. 
The circles, triangles, and crosses denote the thickness of the organic layers and 
are 0.39 [$\mu$m] , 0.95 [$\mu$m], and 1.8 [$\mu$m], respectively. 
The lines indicate the fit by Eq. (\ref{eq:V_resultasympt2}) for the given thickness of the organic layer.  
}
\label{fig:SCL_exp}
\end{figure}


In the conventional method, the current is expressed as a function of voltage to analyze experimental data using the Mott-Gurney equation. 
In order to analyze experimental data using the approximate current-voltage relation given by Eq. (\ref{eq:V_resultasympt2}), 
it is more convenient to express voltage as a function of current as shown in Fig. \ref{fig:SCL_exp}. 
In the figure, the experimental data of Au/$\alpha$-sexithienyl/Au structures are presented in this way. 
The values of mobility obtained by fitting to Eq. (\ref{eq:V_resultasympt2}) were  
$0.0078$ (0.012) [cm$^2$/(Vs)],  $0.015$ (0.013) [cm$^2$/(Vs)], 
$0.025$ (0.027) [cm$^2$/(Vs)] for the sample thickness of 
$0.39$ [$\mu$m], $0.95$ [$\mu$m], and $1.8$ [$\mu$m], respectively. 
The values were close to those in the parenthesis obtained using the Mott-Gurney equation at high voltages in ref. \cite{Horowitz90}. 
We also tried to fit the experimental data by regarding the sample thickness as a free parameter. 
The lines obtained from the fit overlapped with those in Fig. \ref{fig:SCL_exp}. 
Both the mobility and the sample thickness were close to those measured. 
For example, we obtained  $0.0096$ (0.012) [cm$^2$/(Vs)] and the sample thickness $4.1$ [$\mu$m] for 
the data of sample thickness $0.39$ [$\mu$m]. 
These results suggest the validity of Eq. (\ref{eq:V_resultasympt2}) for analyzing the space-charge limited currents. 
When the current-voltage relation was analyzed by using the Mott-Gurney equation, 
the reliable results were obtained by examining thickness dependent of the current given by $I \propto 1/L^3$ in addition to 
the current-voltage relation given by $I \propto V^2$. 
The procedure requires preparation of samples with various thickness keeping the mobility unaltered. 
In contrast, suppose that the current-voltage relation measured for wide range of voltage can be fitted using Eq. (\ref{eq:V_resultasympt2}).  
If the mobility is consistent with that obtained using the Mott-Gurney equation at high voltages, 
and the layer thickness is consistent with that  directly measured, 
the measured currents can be interpreted as space charge limited. 
By using Eq. (\ref{eq:V_resultasympt2}), 
the current-voltage relation can be regarded as space-charge limited currents without examining layer thickness dependence.

In ref. \cite{Horowitz90}, 
the intercept voltage where the linear relation crossed the quadratic relation was independent of the 
thickness of the transport layers for thin layers. 
The intercept voltage divided by $k_{\rm B} T$ was almost independent of temperature for 
temperature above 240 [K].  
These results are again consistent with those in Fig. \ref{fig:SCL_IV_8200}.

\section{Summary and discussion}
\label{sec:summary}

We examined the effect of diffusive currents under the space charge limited condition. 
The Mott-Gurney equation is applicable at high voltages above the onset voltage. 
The current can be fitted by assuming a linear voltage dependence below the onset voltage. 
The onset voltage is around $2.0$ [V] given by $eV/(k_{\rm B} T)=80$ and  
is independent of the mobility and thickness of carrier transport layers. 
Although the current can be phenomenologically fitted by a linear voltage dependence below the onset voltage, 
the actual dependence is very complicated and should be distinguished from Ohm's law. 
We obtained 
an approximate equation which is applicable for voltages satisfying $eV/(k_{\rm B} T)>5$. 
The approximate expression given by Eq. (\ref{eq:V_resultasympt2}) reduces to the Mott-Gurney equation at high voltages. 
Equation (\ref{eq:V_resultasympt2}) approximate the exact numerical current-voltage relation below the onset voltage 
and is applicable above the limit of low voltage around $0.1$ [V]. 
By analyzing experimental results in ref. \cite{Horowitz90} using  Eq. (\ref{eq:V_resultasympt2}), 
both the mobility and the layer thickness were simultaneously obtained 
and the values were consistent with those directly measured in ref. \cite{Horowitz90} 
for thin layers. 

Under the space charge condition, 
carriers are accumulated at the injection side and form the virtual electrode characterized by 
the extremum in the electrostatic potential as shown in Fig. \ref{fig:potential_X}. 
The direction of current flow and that of electric field coincide in the region 
between the virtual electrode and the counter electrode. 
The Mott-Gurney equation is valid when the current is large enough so that the virtual electrode is close to the injection interface. 
As the virtual electrode moves toward the counter electrode by the diffusion effect to homogenize the carrier distribution, 
the current-voltage relation deviates from the Mott-Gurney equation.  

Previously, 
the effect of diffusion on the space charge limited currents was investigated by using the formally exact solutions of the drift-diffusion equation given by 
Eq. (\ref{eq:J}). \cite{Wright61,Sinharay64}
In ref. \cite{Wright61}, 
the reduction of the effective thickness by the factor given by $3/(2^{1/3} z_L)$ was suggested.  
The reduction was attributed to the formation of the virtual electrode where carriers flow from the space charge reservoir.  
In this paper, the location of the virtual electrode is approximately given by Eq. (\ref{eq:virtuale1nd}). 
Although there is a difference in the numerical factor, 
both results are essentially equivalent and share the same scaling law that 
the factor expressing the reduction of the effective thickness scales with $[D/(4 \pi J r_c)]^{1/3}$. 

Even though the exact solutions were obtained, 
the nonlinear equations to determine 
the integration constants were hard to solve analytically. \cite{Wright61,Sinharay64} 
The nonlinear equations were obtained from boundary conditions.  \cite{Wright61,Sinharay64} 
The boundary conditions were evaluated approximately in the article by Wright. \cite{Wright61}
The approximation given by Eq. (39) in ref. \cite{Wright61} can be shown to be essentially equal to 
$C_E \approx a_1/z_L$ obtained from Eq. (\ref{eq:CE_approx}) by noticing $|a_1| 2^{1/3} = 2.946$. 
Strictly speaking, the boundary condition at the counter electrode in this paper is different from that of ref. \cite{Wright61}. 
But we share the conclusion 
that if the approximate boundary condition expressed by $C_E \approx a_1/z_L$ is applicable, 
the current -voltage relation is independent of the boundary condition at the counter electrode. 
Some additional arguments will be given later. 
It should be reminded that the approximate values of the integration constants are used as the seeds 
for evaluating the numerically exact values in this paper. 
Moreover, 
the approximate expression including the diffusion effect is obtained in Eq. (\ref{eq:V_resultasympt2}). 
This expression is simpler than the approximate expression given by Eq. (43) of ref. \cite{Wright61} 
and is tested against the numerically exact results.

In general, 
the boundary condition at the counter electrode can be expressed as,  
$
J =k_e \left(n(L)-n_{\rm BC} (L) \right)
$
. 
By using the boundary condition and Eq. (\ref{eq:sol14}), 
Eq. (\ref{eq:BCz1}) can be generalized as, 
\begin{align}
-\sqrt{\frac{z_L}{\bar{k}_e}+\frac{2 \pi n_{\rm BC} (L) r_c L^2}{z_L^2} + z_1}= f (z_1).   
\label{eq:BCz1_r}
\end{align}
When $z_L/\bar{k}_e+2 \pi n_{\rm BC} (L) r_c L^2/z_L^2 $ is larger than 
$|a_1 |=2.34 \cdots$,  
we can show that 
the discussion considered in analyzing Eq. (\ref{eq:C_B}) holds by using $y_{\rm ik}$ 
and Eq. (\ref{eq:diffC_B_nk}) where $z_L/\bar{k}_e+2 \pi n_{\rm BC} (L) r_c L^2/z_L^2$ is replaced for $z_L/\bar{k}_e$.  
Even by including $n_{\rm BC} (L)$, 
$|C_B| \gg 1$ is satisfied when $z_1\geq 1$. 
In this case, 
$C_B\gg 1$ is satisfied and $C_E$ is approximately given by Eq. (\ref{eq:CE_approx}).  
All the results obtained by taking the limit of $C_B\gg 1$ are not altered. 
The carrier accumulation length given by Eq. (\ref{eq:cacumulation}) and the location of the virtual electrode denoted by $x^\dagger$ 
are not affected by including $n_{\rm BC} (L)$. 
Under the general boundary condition, we still obtain 
the current-voltage relation,  the electrostatic potential, the electric field, the carrier density, 
given by  Eqs. (\ref{eq:V_resultasympt2}), (\ref{eq:pot_approx}), (\ref{eq:solAiry_approx}) and Eq. (\ref{eq:population_approx}).  

The boundary condition of $C_E \approx a_1/z_L$ is later reconsidered in ref. \cite{Sinharay64}. 
In ref. \cite{Sinharay64}, 
Airy functions of real argument were used as fundamental solutions of one-dimensional drift-diffusion equation 
while Bessel functions were used in ref. \cite{Wright61}. 
Since different kinds of Bessel functions were needed at the injection interface and at the boundary of the counter electrode, 
Airy functions are simpler to set boundary conditions 
although Airy functions can be equivalently expressed by using the Bessel functions. \cite{Wright61,Skinner}
As far as we studied using Airy functions, 
the boundary condition given by $C_E \approx a_1/z_L$ 
studied in ref.  \cite{Wright61} may be appropriate for wide range of the currents 
in contrast to the criticism raised in ref. \cite{Sinharay64}. 
The condition given by Eq. (12) of ref. \cite{Sinharay64} equals to setting the boundary condition
$\bar{E} (z_0)=0$ at the injection interface $X=0$. 
The $z$-value satisfying $\bar{E} (z)=0$ is the location of the virtual electrode. 
The approximation of $\bar{E} (z_0)=0$ used in ref. \cite{Sinharay64} corresponds to set the location of the virtual electrode 
being equal to the injection interface. 
The approximation becomes worse as the current decreases. 

One-dimensional drift-diffusion equation was also solved numerically. 
In ref. \cite{Bonham}, 
it was shown that 
the Mott-Gurney equation (Child's law) accurately reproduced the numerical results of current-voltage relation above $5$ V. 
The result and the shape of current-voltage relation are consistent with that shown in Fig. \ref{fig:SCL_IV_8200}. 
At low voltages the current-voltage relation is approximately represented by a linear relation in Fig. \ref{fig:SCL_IV_8200} 
but the current in the intermediate voltage regime is different from both linear and quadratic voltage dependence
as observed experimentally. \cite{Horowitz90} 
By using analytical approach, 
we obtain an approximate current-voltage relation given by Eq. (\ref{eq:V_resultasympt2}). 
The approximate current-voltage relation is applicable even in the intermediate voltage regime.



\renewcommand{\theequation}{A.\arabic{equation}}  
\setcounter{equation}{0}  
\section*{Appendix A. Derivation of Airy function solutions to Eqs. (\ref{eq:Poisson})-(\ref{eq:closedE})}
In order to solve Eq. (\ref{eq:closedE}), it is convenient to introduce dimensionless variables, 
$X\equiv x/L$, $\bar{E} = e E(x) L/(k_B T)$, and 
the dimensionless flux given by 
\begin{align}
J_{dl}= 4 \pi L^3 r_c \frac{J}{D} =2 z_L^3 ,
\label{eq:nondimJe}
\end{align}
where 
 $r_c=e^2/(4 \pi \epsilon \epsilon_0 k_B T)$ is the Onsager length (Coulomb radius). 
 The dimensionless flux, $J_{dl}$ , is related to the dimensionless parameter characterizing the space charge denoted by 
 $z_L$ in Eq. (\ref{eq:zL}).
Equation (\ref{eq:closedE}) can be expressed as, 
\begin{align}
J_{dl} = \frac{\partial}{\partial X} 
\left[ -\frac{\partial}{\partial X} \bar{E} (X) + 
\frac{1}{2} \bar{E}^2 (X)
\right] . 
\label{eq:nondimeclose}
\end{align}
By integrating Eq. (\ref{eq:nondimeclose}), 
we find, 
\begin{align}
J_{dl} \left(X+ C_E\right)= -\frac{\partial}{\partial X} \bar{E} (X) + 
\frac{1}{2} \bar{E}^2 (X), 
\label{eq:nondim_int}
\end{align}
where $C_E$ is a constant to be determined by the boundary condition. 

By introducing $B(X)=-1/\bar{E} (X)$, 
Eq. (\ref{eq:nondim_int}) can be expressed as, 
\begin{align}
J_{dl} \left(X+ C_E\right)B(X)^2= -\frac{\partial}{\partial X} B (X) + 
\frac{1}{2} . 
\label{eq:nondim_B}
\end{align}
By further introducing a new variable given by, 
\begin{align}
z= \left( \frac{J_{dl}}{2} \right)^{1/3} \left( X+C_E
\right), 
\label{eq:sol2}
\end{align}
and transformation, 
\begin{align}
D(X) = 2\left( \frac{J_{dl}}{2} \right)^{1/3} B\left( X \right) ,
\label{eq:sol1}
\end{align}
Eq. (\ref{eq:nondim_B}) can be simplified as, 
\begin{align}
\frac{\partial D}{\partial z} = -D^2 z +1 . 
\label{eq:sol3}
\end{align}
The solution is given by, 
\begin{align}
D(X) =  \frac{h_A (z) }{h_A (z) ^\prime} , 
\label{eq: sol4}
\end{align}
where $h_A (z)$ obeys the Airy equation, \cite{Abramowitz} 
\begin{align}
\frac{\partial^2 h_A (z)}{\partial z^2} - z h_A (z) =0,  
\label{eq:sol5}
\end{align}
and $h_A (z) ^\prime$ denotes the derivative of $h_A(z)$ with respect to $z$. 
Using 
a pair of linearly independent solution of the Airy equation, $Bi (z)$ and $Ai (z)$, \cite{Abramowitz} 
$h(X)$ can be expressed as, 
\begin{align}
h(X) = Bi (z) + Ai  (z)\, C_B . 
\label{eq:sol6_0}
\end{align}
$\bar{E}(X)$ can be expressed using Eq. (\ref{eq: sol4}) and Eq. (\ref{eq:sol6_0}) as, \cite{Sinharay64,Fan48}
\begin{align}
\bar{E} (X) &= -2 z_L \frac{Bi^\prime (z)  + Ai^\prime (z) \,  C_B}{Bi (z) + Ai  (z)\, C_B}  ,
\label{eq:sol13}
\end{align}
where $2^{2/3} J_{dl}^{1/3}=2 z_L$ is used and 
$C_B$ is a constant. 
$Bi^\prime (z)$ and $Ai^\prime (z)$ denote the derivative of $Bi (z)$ and $Ai (z)$ with respect to $z$, respectively. 
Equation (\ref{eq:solAiry}) is obtained by rewriting Eq. (\ref{eq:sol13}). 
$z$ in Eq. (\ref{eq:z}) is obtained by rewriting Eq. (\ref{eq:sol2}) using Eq. (\ref{eq:nondimJe}).

\renewcommand{\theequation}{B.\arabic{equation}}  
\setcounter{equation}{0}  
\section*{Appendix B. Determination of integration constants in Airy functions}

\begin{figure}
\centerline{
\includegraphics[width=1\columnwidth]{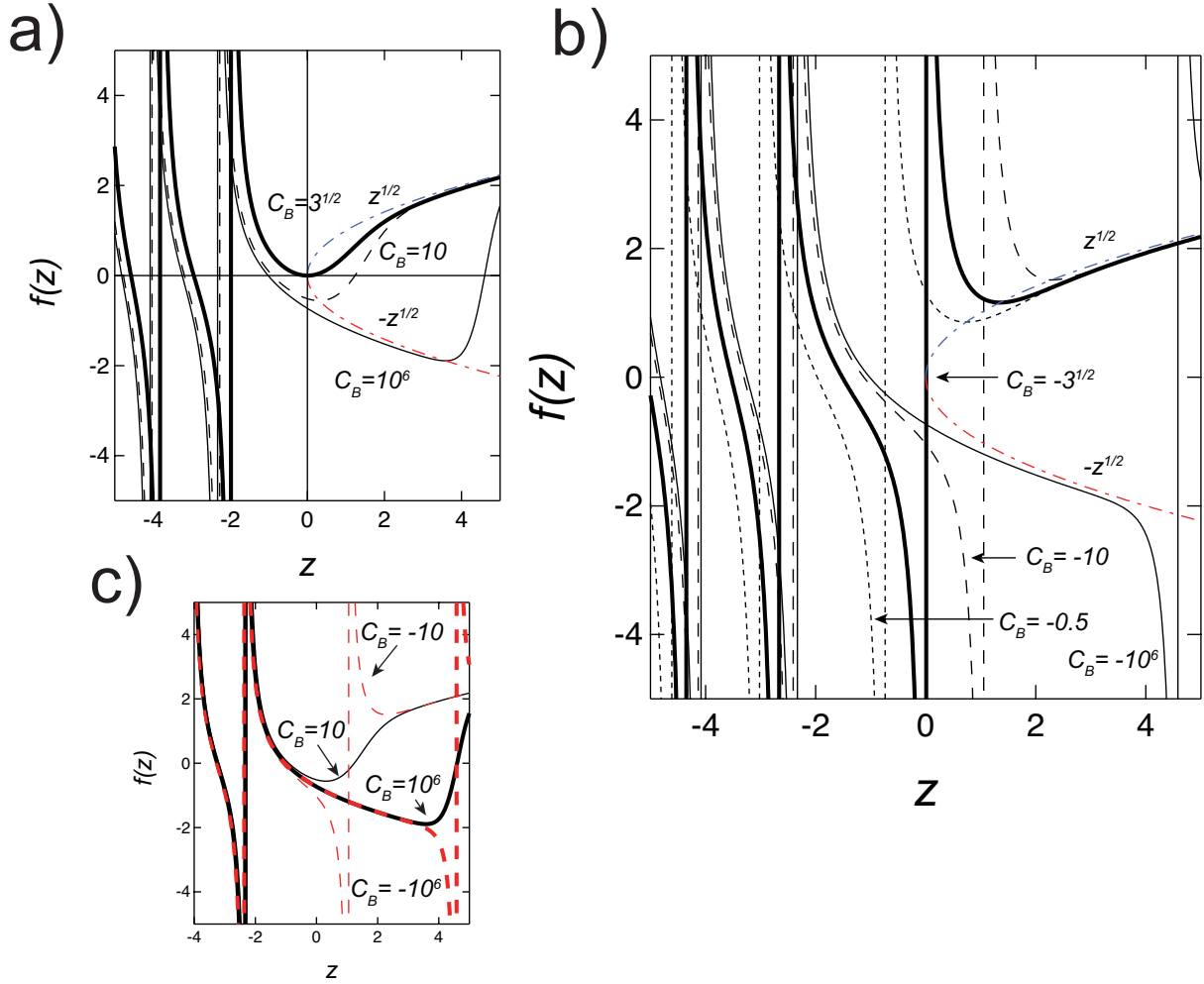}
}
\caption{(Color online) 
$f(z)=-\bar{E} /(2z_L)$, plotted against the variable $z$. 
$z=z_L c_E$ can be used to set the boundary condition at $x=0$ and 
$z=z_L (c_E+1)$ can be used to set the boundary condition at  $x=L$.  
a) $C_B \geq \sqrt{3}$ are used to draw all lines. The thick solid lines, thin solid lines, and  dashed lines indicate 
$C_B = \sqrt{3}$, $C_B = 10^6$ and $C_B = 10$, respectively. 
The (red) dash-dot line below z-axis indicates the line of $-\sqrt{z}$ and the (green) dash-dot line above z-axis indicates the line of $\sqrt{z}$.  
b) $C_B < \sqrt{3}$ are used to draw all lines. The thick solid lines, thin solid lines, dashed lines and dots indicate 
$C_B = -\sqrt{3}$, $C_B = -10^6$, $C_B = -10$ and $C_B = -0.5$, respectively. 
The (red) dash-dot line below z-axis indicates the line of $-\sqrt{z}$ and the (green) dash-dot line above z-axis indicates the line of $\sqrt{z}$.  
c) The thick solid line indicates $C_B = 10^6$ and the thick (red) dashed line indicates $C_B = -10^6$. 
The thin solid line indicates $C_B = 10$ and the thin (red) dashed line indicates $C_B = -10$.
}
\label{fig:SCL_E}
\end{figure}

In Fig. \ref{fig:SCL_E}, 
$f(z)$ in Eq. (\ref{eq:nonE}) is shown for various values of $C_B$. 
$f(z)$ exhibits almost periodic divergent behavior when $z$ is negative. 
The divergence is given by the zero of the denominator of $f(z)$, $Bi (z) + Ai  (z)\, C_B$. 
When $C_B = \sqrt{3}$, 
$f(z)$ touches the x-axis at $z=0$ and never becomes negative for $z>0$. 
When  $C_B > \sqrt{3}$, 
$f(z)$ crosses zero and is negative at $z=0$ and in a region of $z>0$ as shown in Fig. \ref{fig:SCL_E} a). 
As explained below  Eq. (\ref{eq:z0z1}), 
$f (z)$ should be negative at $z=z_1$ and positive at $z=z_0$ and the difference between $z_1$ and $z_0$ is $z_L$. 
$z_L$ defined by Eq. (\ref{eq:zL}) is proportional to the current density. 
When $z$ is smaller than the first zero of $f(z)$ on the negative $z$-axis, $f(z)$ crosses zero almost periodically 
in the middle regions of the sequence of divergent points. 
Obviously, the large difference between $z_1$ and $z_0$ is not possible if $f(z)$ is periodic. 
The large difference between $z_1$ and $z_0$ can be taken for the $z$ values larger than the first zero of $f(z)$ on the negative $z$-axis. 
We consider $f(z)$ in this region when  $C_B \geq \sqrt{3}$. 
Similarly, when $C_B < \sqrt{3}$, the large difference between $z_1$ and $z_0$ can be taken for the $z$ values 
between the largest divergent point and the subsequent largest divergent point as shown in Fig. \ref{fig:SCL_E} b). 
We consider $f(z)$ in this region when  $C_B < \sqrt{3}$. 
As shown in Fig. \ref{fig:SCL_E} c),  
the curve drawn by $f(z)$ using the negative value of $C_B$ coincides with that using the absolute value 
except around the largest divergent point when $C_B < \sqrt{3}$. 
$f(z_0)$ should be positive and is not close to the largest divergent point, where $f(z)$ becomes negative. 
It implies that the value of $f(z_0)$ will not be influenced by changing the sign of $C_B$.  

In Fig. \ref{fig:SCL_E} a)-b), we plotted $\sqrt{z}$ as well as $f(z)$. 
When $\bar{n}(0) > 10$, the left-hand side of Eq. (\ref{eq:BCz0}) becomes large.  
$z_0$ obtained from the crossing point between $\sqrt{\bar{n}(0)/(2 z_L^2)+ z}$ and $f(z)$ is close to $-2$ 
when $|C_B| \gg 1$ as shown in Fig. \ref{fig:SCL_E}. 
Under the condition of $|C_B| \gg 1$, 
the integration constant $C_E$ can be determined from $z_0$ by using both Eq. (\ref{eq:BCz0}) and approximate expression of 
$f(z)$, 
\begin{align}
f_a (z) \approx \frac{ Ai^\prime (z) }{Ai  (z)}   . 
\label{eq:nonEa}
\end{align}
The divergence of $f (z)$ around $z=-2$ is approximately obtained from the first zero of $Ai  (z)$ in the negative $z$-axis given by 
\begin{align}
a_1\approx - 2.34. 
\label{eq:a1}
\end{align}
By approximating the crossing point between 
$\sqrt{\frac{\bar{n}(0)}{2 z_L^2}+ z_0}$ and $f (z_0)$ by $a_1$, 
the left-hand side of Eq. (\ref{eq:BCz0}) can be approximated as 
$\sqrt{\bar{n}(0)/(2 z_L^2)+ a_1}$. 
The approximate value of $C_E$ can be obtained by further introducing the expansion around $a_1$, 
\begin{align}
f_a (z) \approx \frac{1}{z_L C_E - a_1}  ,
\label{eq:nonEa_approx}
\end{align}
into Eq. (\ref{eq:BCz0}) 
as
\begin{align}
C_E \approx \frac{1}{z_L} \left( a_1 + 
\frac{1}{\sqrt{\bar{n}(0)/(2 z_L^2)+ a_1}}
\right). 
\label{eq:CE_approx_Ap}
\end{align}
In this way, 
Eq. (\ref{eq:CE_approx}) was derived. 
The integration constant $C_E$ is determined from the boundary condition representing the space-charge injection of carriers 
under the assumption of $|C_B| \gg 1$. 

In the rest of the Appendix B, 
we study the lower bound of $|C_B|$ using the boundary condition at the counter electrode. 
The boundary condition representing the fast extraction of carriers to the counter electrode with the rate $k_e$ is given by 
$
J =k_e n(L)
$. 
By using the boundary condition and Eq. (\ref{eq:sol14}), 
we obtain, 
\begin{align}
-\sqrt{\frac{z_L}{\bar{k}_e}+ z_1}= f (z_1),  
\label{eq:BCz1}
\end{align}
where the dimensionless extraction rate is defined by $\bar{k}_e=k L/D$. 
The minus sign in Eq. (\ref{eq:BCz1}) indicates the positive electric field by using Eq. (\ref{eq:nonE}). 
We note using  Fig. \ref{fig:SCL_E} that Eq. (\ref{eq:BCz1}) has a solution 
in the region considered for $z_1$ only if $C_B$ is negative when $z_L/\bar{k}_e$ is smaller than 
$a_1=-2.34 \cdots$. 

We obtain from Eq. (\ref{eq:BCz1}), 
\begin{align}
C_B = - \frac{\sqrt{\left(z_L/\bar{k}_e\right)+z_1}{\rm Bi} \left(z_1\right)+{\rm Bi}^\prime \left(z_1\right)
}{\sqrt{\left(z_L/\bar{k}_e\right)+z_1}{\rm Ai} \left(z_1\right)+{\rm Ai}^\prime \left(z_1\right) 
} .
\label{eq:C_B}
\end{align}
By introducing a new variable $y_{\rm ik}=z_L/\bar{k}_e$ and differentiating $C_B$ with respect to $y_{\rm ik}$ we find 
that the derivative is never negative as shown below 
\begin{align}
\frac{1}{2 \pi \sqrt{\left(z_L/\bar{k}_e\right)+z_1}
\left[ \sqrt{\left(z_L/\bar{k}_e\right)+z_1}{\rm Ai} \left(z_1\right)+{\rm Ai}^\prime \left(z_1\right) \right]^2
} .
\label{eq:diffC_B_nk}
\end{align}
When $C_B$ is negative and increases with increasing $y_{\rm ik}$, we obtain the smallest $|C_B|$ by taking the infinite limit of $y_{\rm ik}$ in
$C_B$ as 
$-{\rm Bi} \left(z_1\right)/{\rm Ai} \left(z_1\right)$. 
$|{\rm Bi} \left(z_1\right)/{\rm Ai} \left(z_1\right)|$ increases with increasing $z_1$ and the smallest value is given at $z_1=0$ as $\sqrt{3}$. 
The similar consideration leads to $\sqrt{3}$ as the smallest value of $|C_B|$ when $C_B$ is positive. 
Therefore, $|C_B |> \sqrt{3}$ is obtained.

\renewcommand{\theequation}{C.\arabic{equation}}  
\setcounter{equation}{0}  
\section*{Appendix C. Current-voltage relation when $E(0)=0$}

We consider the boundary conditions expressed as, 
\begin{align}
E(0) &=0,
\label{eqE:BC1}\\
n(L) &=0. 
\label{eqE:BC2}
\end{align}
The first boundary condition given by Eq. (\ref{eqE:BC1}) 
leads to an equation to determine the integration constant $C_E$, 
\begin{align}
{\rm Bi}^\prime \left(z_L C_E\right)+ C_B {\rm Ai}^\prime \left(z_L C_E\right) =0. 
\label{eqE:C_E}
\end{align}
When the electric field is positive $E(x) >0$, 
the boundary condition Eq. (\ref{eqE:BC2}) together with Eq. (\ref{eq:comb}) leads to  
\begin{align}
C_B = - \frac{\sqrt{z_L\left(1+C_E\right)}{\rm Bi} \left[z_L\left(1+C_E\right)\right]+{\rm Bi}^\prime \left[z_L\left(1+C_E\right)\right]
}{\sqrt{z_L\left(1+C_E\right)}{\rm Ai} \left[z_L\left(1+C_E\right)\right]+{\rm Ai}^\prime \left[z_L\left(1+C_E\right)\right]
} .
\label{eqE:C_B_C_E}
\end{align}
By substituting Eq. (\ref{eqE:C_B_C_E}) into Eq. (\ref{eqE:C_E}), we obtain an implicit function of $C_E$. 
$C_B$ can be determined from Eq. (\ref{eqE:C_B_C_E}) using the value of $C_E$.    

First, we consider the case of  $z_L\ll 1$.
By substituting Eq. (\ref{eqE:C_B_C_E}) into Eq. (\ref{eq:V_formal}) and 
expanding the right-hand side of Eq. (\ref{eq:V_formal}) in terms of $z_L$, 
we obtain 
\begin{align}
\frac{eV}{k_{\rm B} T} \approx 2 \sqrt{(1+C_E) z_L^3}-\frac{1}{3}z_L^3+\frac{1}{6}   \sqrt{(1+C_E) z_L^9} +\cdots . 
\label{eqE:Vexpand1}
\end{align} 

By expanding $C_B$ in terms of $1+C_E$, we obtain, 
\begin{align}
C_B \approx \sqrt{3} + \frac{2 \Gamma(1/3)^2}{3^{1/6} \Gamma(2/3)^2} (1+C_E) z_L .
\label{eqE:expansionC_B_a1}
\end{align}
By substituting Eq. (\ref{eqE:expansionC_B_a1}) into Eq. (\ref{eqE:C_E}), we obtain $C_E\approx -1$ by 
noticing 
${\rm Bi}^\prime (0)+\sqrt{3} {\rm Ai}^\prime (0)=0$. 
Further expansion of ${\rm Bi}^\prime \left(z_L C_E\right)+ C_B {\rm Ai}^\prime \left(z_L C_E\right) $ in terms of $1 + C_E$ and $z_L$  
yields 
$
C_E \approx -1 + z_L^3/4 
$. 
By substituting the above expression of $C_E$ into Eq. (\ref{eqE:Vexpand1}), we obtain 
$eV /(k_{\rm B} T)\approx (2/3) z_L^3$ which can be expressed as  
\begin{align}
I = 3\epsilon \epsilon_0 \mu k_{\rm B} T \frac{V}{L^3},  
\label{eqE:lowfield}
\end{align} 
using the Einstein relation $D=\mu k_{\rm B} T$. 
The current density shows $1/L^3$-dependence exactly in the same way as 
the Mott-Gurney equation. 
The conductivity $\sigma$ is obtained from $I/(V/L)$ as 
\begin{align}
\sigma= \frac{3\epsilon \epsilon_0 k_{\rm B} T}{L^2} \mu =3 e^2 \frac{1}{r_c L^2} \mu,  
\label{eqE:conductivity_low}
\end{align} 
where $e \mu$ is the electrical mobility defined before. 
The equation similar to Eq. (\ref{eqE:lowfield}), $I=2 \pi^2 \epsilon \epsilon_0 \mu k_{\rm B} T V/L^3$, 
was obtained previously. \cite{Koehler}

In the case of $z_L\gg 1$, 
we note that 
$C_B$ in Eq. (\ref{eqE:C_B_C_E}) rapidly increases as $z_L$ increases by using numerical evaluation. 
The growth is faster than that estimated by using 
Eq. (\ref{eqE:expansionC_B_a1}) and we take the limit of $C_B \gg 1$ in Eq. (\ref{eqE:C_E}). 
In this limit, $C_E$ can be obtained from 
${\rm Ai}^\prime \left(z_L C_E\right) =0$. 
Using the first zero on the negative real axis,  we have, \cite{Abramowitz}
$
C_E \approx - \left( 3 \pi \right)^{2/3}/(4 z_L)$ .
By using the same approximation leading to Eq. (\ref{eq:Child}), we again obtain 
the Mott--Gurney equation when $z_L\gg 1$. 
\begin{figure}
\centerline{
\includegraphics[width=0.6\columnwidth]{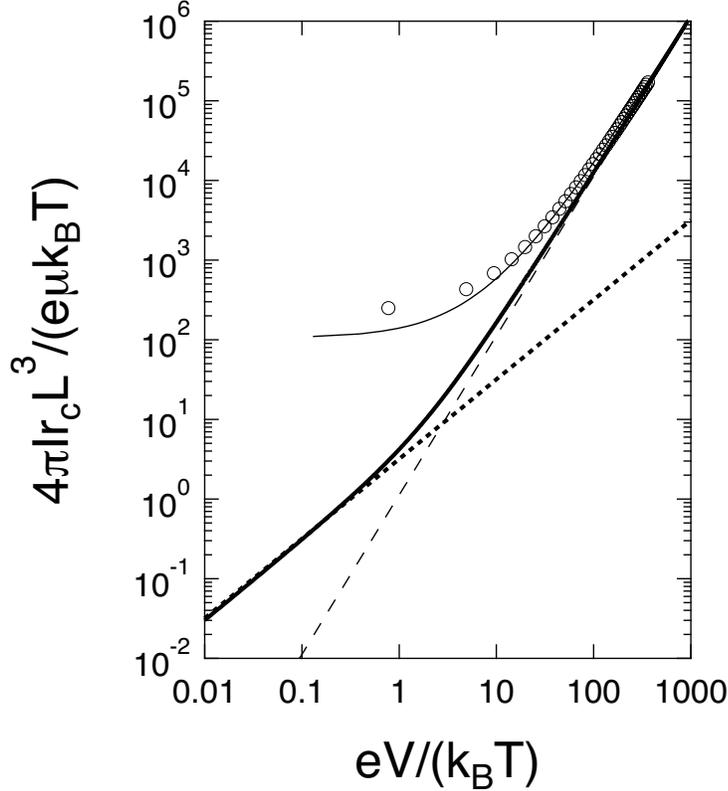}
}
\caption{Dimensionless current  
$4 \pi I r_c L^3/(e \mu k_{\rm B} T)$ as a function of 
dimensionless voltage $eV/(k_{\rm B} T)$. 
The thick solid line represents the exact numerical solution of Eq. (\ref{eq:V_formal}) using the boundary condition,  
$E(0)=0$.
The dotted line represents the result of Eq. (\ref{eqE:lowfield}). 
The circles represent the solution of Eq. (\ref{eq:V_formal}) using the boundary condition,  
$\bar{n} (0) = 8.2 \times 10^4$ and (\ref{eqE:BC2}). 
The thin solid line and dashed line represent the results of Eq. (\ref{eq:V_resultasympt2}), 
and the Mott--Gurney equation, Eq. (\ref{eq:Child}), respectively.
}
\label{fig:SCL_comp}
\end{figure}


The intercept current density $I^*$ between linear and quadratic regime of the current-voltage relation 
is obtained from the condition $z_L=1$ as, 
\begin{align}
I^*  =\frac{2 \mu \epsilon \epsilon_0 \left(k_{\rm B} T \right)^2}{e} \frac{1}{L^3}. 
\label{eq:inflectionI}
\end{align}
The intercept current density is related to the thickness of carrier transport layers by $1/L^3$-dependence; 
the intercept current density increases rapidly by decreasing the thickness of carrier transport layers. 
The intercept voltage $V^*$ can be well approximated by  
\begin{align}
\frac{eV^*}{k_{\rm B} T} =1, 
\label{eq:inflectionB}
\end{align}
judging from Fig. \ref{fig:SCL_comp}. 
The intercept voltage is approximately estimated from the thermal energy as $0.026$[V] regardless of the thickness of the carrier transport layers. 
The value of the intercept voltage is smaller than that obtained by taking into account the shift of the virtual electrode due to the diffusion effect.


\end{document}